# Brewing Analytics Quality
# For Cloud Performance

Li Chen[1], Pooja Jain[2], Kingsum Chow[1], Emad Guirguis[1], and Tony Wu[1]

{li.chen, kingsum.chow, emad.guirguis, tony.wu}@intel.com, pxj131230@utdallas.edu

## Abstract

Cloud computing has become increasingly popular. Many options of cloud deployments are available. Testing cloud performance would enable us to choose a cloud deployment based on the requirements. In this paper, we present an innovative process, implemented in software, to allow us to assess the quality of the cloud performance data. The process combines performance data from multiple machines, spanning across user experience data, workload performance metrics, and readily available system performance data. Furthermore, we discuss the major challenges of bringing raw data into tidy data formats in order to enable subsequent analysis, and describe how our process has several layers of assessment to validate the quality of the data processing procedure. We present a case study to demonstrate the effectiveness of our proposed process, and conclude our paper with several future research directions worth investigating.

## Biography

*Li Chen is a data scientist at Intel. She received her PhD degree in Applied Mathematics and Statistics from the Johns Hopkins University in 2015. She specializes in applying analytics to system performance data.*

*Kingsum Chow is a principal engineer at Intel. He received his PhD in Computer Science and Engineering from the University of Washington in 1996. He specializes in performance, modeling and analysis of software applications, and leads analysis of cloud application performance. In his spare time, he volunteers to coach multiple robotics teams to bring the joy of learning Science, Technology, Engineering and Mathematics to the students.*

*Pooja Jain is pursuing her MS degree in Computer Science and Engineering at University of Texas, Dallas. She is an intern at Intel. She specializes in characterizing user behavior for cloud applications.*

*Emad Guirguis is a software performance engineer at Intel. He received his MS degree in Computer Science from Texas State University-San Marcos in 2012. He specializes in optimizing software applications in the cloud.*

*Tony Wu is a system architect and an engineering manager at Intel. He received his PhD degree in Electrical Engineering from the University of Minnesota-Twin Cities in 2005. He specializes in optimizing software applications.*

---

[1]    System Technologies and Optimization, Software and Services Group, Intel Corporation
[2]    Computer Science and Engineering, University of  Texas at Dallas, TX

# 1   Introduction

According to the definition of Cloud Computing by the National Institute of Standards and Technology (NIST), "Cloud computing is a model of enabling ubiquitous, convenient, on-demand network access to a shared pool of configurable computing resources that can be rapidly provisioned and released with minimal management effort or service provider interaction." There are five essential characteristics for the cloud computing model: On-demand self-service, resource pooling, rapid elasticity, measured service, and broad network access. These characteristics are met through three types of service models for cloud computing: Software-as-a-service (SaaS), Platform-as-a-service (PaaS), Infrastructure-as-a-service (IaaS), and four types of deployment models: public cloud, private cloud, community cloud, and hybrid cloud [1]. Such easy to deploy properties are propelling a transition into cloud computing.

Enterprises used to run their applications on dedicated servers and hardly ever on a shared or cloud environment. Once it was apparent that even enterprises were looking to go along the cloud path and spend big in the process, testing the application performance in the cloud has become essential. Java started being featured in some hosting/cloud solution. The interaction of Java garbage collections across applications brings an additional challenge in assessing cloud performance. Cloud application usage is very variable, and there are several ways to measure performance; thus optimization can be done in many directions, where improvement in one factor may negatively impact another. End user satisfaction and quality of service establish the guideline for performance optimization. The complexity of the Java Cloud solution, including several services such as IaaS, PaaS and SaaS, adds to the tediousness of measuring the performance. The main challenges lie in two areas: there are complicated interactions between multiple applications running on the same machine, and different loads may be applied to those applications. Early work on delivering quality in software performance and scalability testing [2] needs to be extended to cloud computing.

In this paper, we describe a process, implemented in software, to assess the quality of cloud performance data. This process combines performance data from multiple machines, spanning across user experience data, workload performance metrics, and readily available system performance data. These performance statistics have been readily used to classifying Enterprise applications [3]. The data collection and processing procedure across the system is able to generate concrete data for posterior analytics. Principled statistical analysis on cloud performance data will serve as valuable tool to assess cloud performance.

User experience data, such as throughput and response time, can be obtained from typical load driver systems used in software testing. One typical load driver, Faban [4], is a driver development framework used in SPECjEnterprise [5], SPECvirt [6] and SPECsip [7] benchmarks, and can control a number of load generation nodes. Such a framework defines operations, transactions and associated statistics collection and reporting. Faban is one of the tools that can be used for cloud computing workloads.

System Activity Report (SAR) [8] on Linux systems can save system performance data such as CPU activity, memory/paging, device load, network, etc. It writes the contents of selected cumulative activity counters in the operating system. The accounting system, based on the values in the count and interval parameters, writes information the specified number of times spaced at the specified intervals in seconds. Performance Counters for Linux (Perf) [9], a Linux profiling tool for performance counters, can add additional detailed performance counters for software processes and microarchitectures.

One challenge of dealing with performance data sets is bringing them into a tidy data format [10] for analysis. Before discussing the issues of bringing performance datasets into tidy data format, we first provide the motivation of why such an action is desired. Cloud workload is much more complex than enterprise workload. Manually examining the cloud performance data may not be the optimal way. In addition, workload modeling requires the dataset usable for statistical modelling purposes. Dynamic analysis on the time series cloud performance data can provide new insights for cloud performance optimization. Techniques such as model selection, non-stationary time series analysis, and stochastic processes can be adopted to analyze the cloud workload. Furthermore one could also borrow the techniques such as spectral clustering [11], sparse representation classification [12] [13], vertex nomination [14] [15] and graph matching [16] [17] in the field of random graph inference to model the user behavior graph of the cloud workload. All these inference frameworks and methodologies would benefit from the tidy data format.

There is no standard format in which the data is collected. Therefore, parsing the data into a simple and coherent format is as essential as collecting and analyzing the data. Data munging gets tedious when dealing with data sets collected from different sources. Merging these files based upon time series is one of the major challenges. Elaborating on this, various data sets might have different formats of time in which the data is collected, namely the a) world clock time and b) epoch (UNIX time). , which might further differ in the following criterion: i) time zones, ii) units of measurements (seconds or milliseconds), iii) the time interval at which the data is collected and iv) the frequency at which data samples are collected. The data set is too huge to manually check consistency of results produced by the software components.

To overcome these challenges, we first add a set of basic queries and scripts to check the correctness of the data and data format. As data can vary a lot, we then add a set of performance models to check the quality of the performance assessment. By using ensembles of performance models, we minimize potential errors in the software performance assessment tool chains.

Our paper has the following three key contributions: firstly, we define the taxonomy of challenges in identifying performance bottlenecks in the cloud; secondly, we develop a software tool that applies analytics to test the application performance in the cloud; and lastly, we establish a methodology to evaluate and improve the quality of the analytics used for cloud performance assessment. We organize our paper as follows: In Section 2, we discuss the motivation for proposing our software for cloud performance assessment and the major challenge behind it. In Section 3, we present in detail our software chain in order to assess cloud performance. In Section 4, we demonstrate in a case study of how our software is used. Section 5 summaries our paper and discusses future direction for this line of work.

## 2 Challenges

Analyzing cloud performance data is one way to assess cloud performance. While we expect insights from advanced statistical analysis or machine learning techniques, the first hurdle is to bring collected performance data into the tidy data format [10] before analytics are applied. As different performance tools generate data in different formats, parsing the data into a simple and coherent format is just as essential as collecting and analyzing the data. Moreover, validating the quality of the data processing procedure is necessary; otherwise all the subsequent work becomes in vain. In this section, we first start with a primitive and idealized scenario for quality assessment for data processing, then define the challenges in identifying performance bottlenecks in the cloud, and present our insights on dealing with and working around these challenges.

### 2.1 An Idealized Scenario: A Tale of Two Engineers

Here, we explore an idealized scenario for the purpose of illustrating how to assess the quality of processing raw performance data. Suppose two engineers each use a different program and write separate scripts to process the raw data. The first engineer exports her processed data into the traditional n-by-m matrix formats, where the rows denote the number of samples (depending on the situation, the samples can be time stamps for example), and the columns denote the dimension of cloud performance metrics. The second engineer also exports her processed data into the traditional tidy data format. Although the two engineers participate in the same process – transforming the same raw data into a traditional data matrix, the resulted processed data could be different due to error in coding. The most straightforward validation on data quality is to compare the two processed datasets entry-by-entry.

## 2.2 Back to Reality: Way More Challenges Ahead

In reality, way more challenges lie ahead to transform the data and validate the quality of processing. The term "data munging" refers to the mapping from raw data into another convenient format useful for further purposes. In our case, this format will conveniently enable data visualization, data aggregation, statistical analysis and predictive modeling.

Merging the several files, which are collected from multiple sources, based upon time series is not trivial. In practice, cloud performance data comes in high volumes, variable velocity, and different raw formats. Hence, manually checking the consistency of the datasets can be exhausting.

Even after multiple datasets are merged with consistent alignment of the time samples, the quality of munging still needs more assessment. The issues often arise when the data transformation method is incorrect, causing inconsistency between the raw dataset and processed & merged datasets. Again, manually checking consistency here is a difficult task, especially during the age of big data.

# 3 Assessing Analytics Quality for Cloud Performance

To overcome these challenges, we developed an approach consisting multiple layers of quality assessments. Our goal is to assess the quality of the processed data for further analysis. In a nutshell, we first add a set of basic queries and scripts to check the correctness. As data can vary a lot, we then add a set of performance models to check the quality of the performance assessment. By using ensembles of performance models, we minimize potential errors in the software performance assessment tool chains. Figure 1. A quality model for Cloud performance analyticsdepicts the flow of our proposed software. In the following subsections, we describe in detail how our software works.

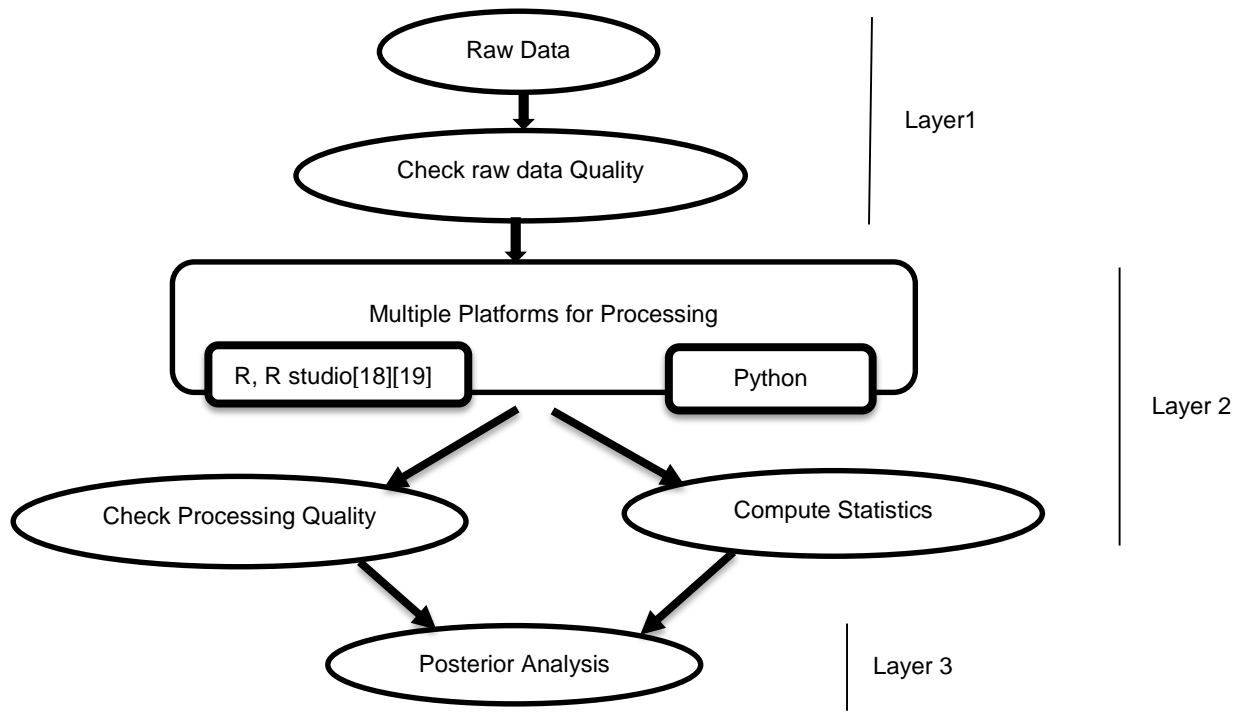

**Figure 1. A quality model for Cloud performance analytics**

## 3.1 Layer 1: Raw Data Quality Assessment

We obtain raw performance data from multiple sources such as SAR, Faban or other cloud testing sources. The very first step is to check the quality of the raw data; otherwise error propagates through the entire processing flow. We note that there are no universal rule for checking the quality of the raw data, as the rule checking strongly depends on the data source. For example, using the Faban driver, one rule to check the quality of the raw data is to examine a combination consisting of a particular time stamp and a particular machine name. This rule check is easily implemented using packages in R [18] or R studio [19], which reshape the data with counter values fitted in the data table.

## 3.2 Layer 2: Processing Quality Assessment

After verifying the quality of the raw data, the next step is to process the raw data into tidy data formats, which are convenient for further performance analysis. Our software model enables multiple programming languages, including Python and R [18], to process the raw data. Furthermore, our software model incorporates the freedom of using various packages even within a particular language. We define this step of our software as "multiple platforms for data processing".

The key design reason that the raw data are processed via different channels is that this gives an extra layer of quality assessment. When the tidy data formats are generated using various package tools in different processing languages, our software can check the consistency of the processed datasets across multiple platforms. We believe this is a necessary step to guarantee data processing quality.

After the consistency of the datasets across multiple platforms is confirmed, we introduce another quality assessment, which is to compute statistics. Stated generally here, this step is also domain-specific. Our software enables such consistency checks by implementing rules based on domain expertise. For instance, the clock-per-instruction (CPI) cannot be lower than 0.25, or the CPU utilization cannot be over 100%. We note that these rules are not the gold standards, because real cloud performance data may have violated some rules, but the data itself is correctly measured.

## 3.3 Layer 3: Posterior Analysis Assessment

The last layer of quality assessment is posterior analysis. After the raw data is formatted into a tidy format, which can be input for statistical software tools such as R, posterior analysis delivers the final layer of quality assurance in our software. We test the posterior consistency by applying principled data analytics methodologies and detect anomalies that exist within the datasets. Although the anomalies may exist due to the intrinsic and unavoidable noise in the dataset, some anomalies directly reflecting the poor quality of the datasets can still be identified using basic statistical methods. Particularly, the statistical tools such as outlier detection, clustering and density fit can pinpoint the "impossibles" in the data. This will alert the engineers to have a second look at the data processing procedure. We note that the step of posterior analysis is important for assuring the quality of the data processing procedure. However, practitioners will not rely solely on this step, since most mistakes should be captured in the earlier two layers of quality assurance.

## 3.4 Test Cases

Although not presented in depicting our approach, we consider several test cases to validate the data processing quality. The key reason for including test cases is that, with prior knowledge of the test cases, we expect to see certain outputs after data processing. When such outputs are missing, we know the data processing procedure contains errors, and does not pass our quality assessment.

# 4  Case Study

In this section, we present a case study of processing cloud performance data using our approach, and demonstrate that our approach, with several layers of quality assessment, is capable of capturing mistakes during data processing.

## 4.1  A Cloud Workload

For our demonstration we use a Software-as-a-Service (SaaS) workload composed of dozens of Java applications serving requests in a group of domains. The details of these applications are beyond the scope of this paper. As illustrated in Figure 2, the workload is driven by five groups of users simulated on the driver. Each user group simulates a particular type of users, sending a sequence of requests to the service. Upon receiving the response to a request, each virtual user waits for a period of time, called the think time, before sending the subsequent request. Different number of virtual users are assigned to each user group. In an experiment, they are gradually increased to a maximum level, based on the design of the experiments.

On the server side simulating the cloud computing, there are 54 Java Virtual Machine (JVM) applications running as the SaaS workload, responsible for addressing different sets of requests. For instance, consider it as an employee portal offering various services for management of wages/vacation tasks etc, and each service handled by set of JVM instances. These Java applications are configured with different parameters. They have different run-time behavior from each other. They also interfere with each other in the cloud. There is also a database process providing data support to these applications.

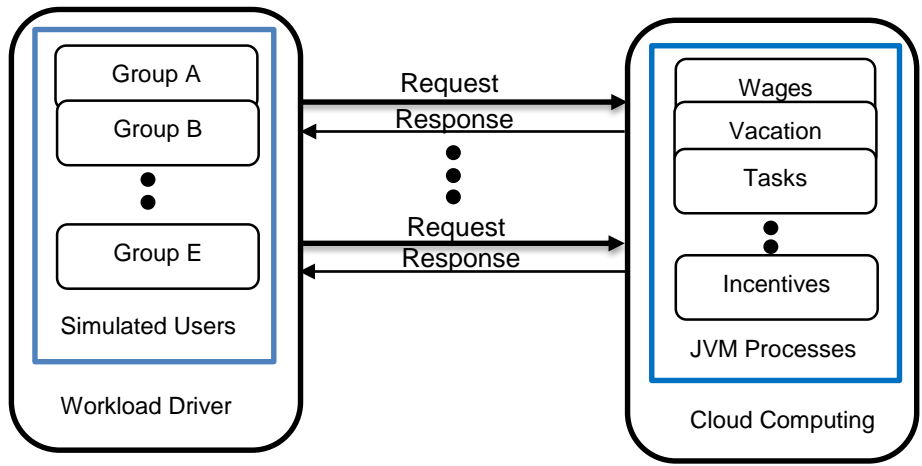

**Figure 2. A Cloud workload is composed of a driver of multiple user groups sending requests to and receiving responses from the services in the cloud.**

## 4.2  Performance Data Collection

We demonstrate our experiment using two machines, one as a web server and the other as a client server. The client server hosts an application driver that generates the workload. As described in the previous section, it drives the process of sending in the requests to the server through virtual users. In parallel it also increases the load on the server by ramping up the number of virtual users interacting with the server. There are 45 application servers running over 54 Java virtual machines, a combination of which are responsible in processing each request.

To learn the performance of the cloud workload, we collect data on both the machines, or say from every source applicable. Therefore we collect data from every domain that reflects performance of the workload, such as on server, or system level, which is determined by operating system. This data usually helps us determine the bottlenecks of the system. That is, it gives us an idea about how much the system resources (CPU, input/output, memory, and network) are being utilized over time. These application specific data points usually describe the behavior of the application, given the load. In our case we collect java garbage collection (GC) logs of each of the 45 application servers to see their activity and interaction among themselves. On the driver simulating virtual users, we collect workload specific data, which emphasizes the user experience. The driver records data like the number of users, response time, failed transactions etc.

As the sources of the data collected vary as described above, the datasets vary from each other in terms of frequency of data points, metrics collected, units of measurements, etc. For example, if we consider frequency, the SAR data is collected on a server every 10 seconds, so that it doesn't impact the system performance, whereas the workload data on a client is collected every 5 seconds, as the performance on client server is not critical. To some extent, the frequency at which the data is collected via these tools is still controllable, but when it comes to the application specific data, in our case the GC logs, the data points are only recorded when a garbage collection takes places. The kinds of metrics collected by each tool is different, but getting all of this data together in a tidy format for analysis may help us generate an insight about which attributes impact the other or how they are correlated.

### 4.3  Performance Data Processing

We describe how we process the data collected from SAR, GC and client server. The original garbage collection data is in log files. The log files are human-readable: for example, each line may state the time when garbage collection occurs, the time elapsed since start of the workload, GC type (either GC or Full GC), memory before and after GC, heap size, and GC pause time. Performance engineers may manually examine the log files line by line to identify performance issues. The purpose of developing our tool is to enable an automated process for cloud performance analytics. Additionally, our tool provides the first necessary step to apply multivariate time series analysis to the performance data. Our tool uses Python scripts to parse the log files line by line. The processed GC dataset has time stamps in epoch milliseconds in each row, and attributes in each column. The time stamps correspond to when GC activity happens. The attributes are memory before GC, memory after GC, heap size, GC type (GC or Full GC) and GC pause time. The resulting data set is thus in tidy data format, and simple statistical analysis implemented in R can be done on this dataset. Furthermore, transforming the GC log files into tidy data formats provides a common ground for merging the SAR data and client server data, and this gives a better perspective on the cloud workload. The role of our tool is to assure the quality of the data is sound.

The second source of data is collected from the System Analysis Report. Recall that SAR records performance every 10 seconds in this workload. Therefore, the file is structured such that the time stamps (in epoch seconds) are in each row with increment of 10 second, and the features are in each column. The features are mainly describing the aspects of CPU, memory, disk I/O and network. Some examples of features include CPU percentage utilization, paging statistics, packets received statistics and so on.

Particularly for this workload, we also have performance data collected from the user experience perspective. As discussed before, this data is collected by a specific client driver, which is a software tool to simulate user groups, where each group has specific user behaviors. The original data format is in terms of unique pairs of time stamps and performance counters, i.e., (T1, P1), where T1 is the recorded time stamp and P1 is the recorded performance counter at T1; one can think of it as edge lists. The data format is different from the GC logs and the SAR data, and provides challenges to directly merge this data with GC and SAR. Moreover multivariate time series analysis cannot directly be done on the paired data, and therefore our tool plays the role of reshaping the data set to enable subsequent merging and analysis. Here we transform the unique pairs into standard data format such that the rows are time stamps (in unit y:m:d:h:m:s) and the columns are performance counters for the user behaviors. Some examples of performance counters are server times, web pages received per second, kilobytes received per second and so on. Note that the time increment is 5 seconds, as the client server records the data every 5 seconds. Data processing on the client data is done using R using specific packages. To guarantee the quality of processing this dataset, our software implements two independent methods to transform the client data to tidy data format, and compares the consistency of the transformed datasets entry-by-entry. This step provides an extra layer to assure the quality of data processing.

### 4.4 Merging of the Performance Datasets

The step of data processing described in Section 4.3 transforms the SAR, GC and client server datasets into three tidy data formats respectively. These three datasets describe the workload from three aspects. Statistical analysis can be done directly and separately on these datasets. However our software aims at providing a tool to bring together all three datasets, so that statistical analysis can be done on the entire unified dataset, and explore the correlation among SAR, GC and client server. There are several challenges for merging these datasets. Firstly, the units of time in these three datasets are different: one in world clock, one in epoch seconds, and one in milliseconds. Secondly, the mean of the time stamps vary across three datasets. The time stamps for GC dataset denote the occurrence of GC activities, and thus the increments between the time stamps are not constant, and but are in fact random. On the other hand, the increments between the time stamps for SAR data are 10 seconds, as the system-level performance counters are recorded and reported every 10 seconds. The increments between the time stamps for the client data are 5 seconds, as the client driver records the response time and user experience related performance metrics every 5 seconds. Our tool overcomes the challenges of time stamp unit and interval discrepancy by converting all time units into epoch milliseconds and interlacing the epoch times of all three datasets. Therefore, our tool is the first tool to enable cloud workload characterization from system, applications and client perspective. This provides a convenient platform for subsequent statistical analysis for cloud workload performance.

### 4.5 Performance Data Profiling and Analytics

After the step of data merging, we have one coherent data set containing multivariate time series to describe the cloud workload. The rows in the dataset are the time stamps in epoch milliseconds and the columns in the dataset are performance metrics for operating system, user experience as well as garbage collection statistics. We have discussed the advantage of utilizing such a merged dataset. At the same time, some issues arise that are worth our attention. First of all, if the original SAR, GC and Client data matrices have dimensions k by l, m by n, and x by y, then the resulted data matrix has dimension a by b, where a ≤ k + m +x, and b = l+n+y. This merged data matrix has more time samples and a higher number of attribute dimensions. Secondly, the merged data matrix has a notable number of missing values, which are created artificially when merging and interlacing the time stamps across three datasets.

Although one could examine the merged data matrix entry by entry, it is more economical and practical to appeal to data profiling on the big data matrix to assess its quality, especially for the reason that all three datasets were examined entry-by-entry respectively as discussed in Section 5.2. Our tool implements the calculations of mean, median, minimum, maximum, range, the number of missing values, and percentiles to profile the datasets. These computations of the profiling statistics are implemented in R and formulated as below. For each performance metric X, let X1, X2, …, XT denote the time series of X during the cloud workload. The software profiles the performance metric by calculating:

$$mean(X) = \frac{1}{T} \sum_{i=1}^{T} X_i.$$

$$median(X) = midpoint(X_1', X_2', \ldots, X_T'), where\ X_1' \leq \cdots \leq X_T'\ and\ \{X_1', X_2', \ldots, X_T'\} = \{X_1, X_2, \ldots, X_T\}.$$

$$minimum = min_i\{X_1, X_2, \ldots, X_T\}.$$

$$maximum = max_i\{X_1, X_2, \ldots, X_T\}.$$

$$range(X) = maximum - minimum.$$

$$Number\ of\ missing\ values\ for\ X = \sum_{i=1}^{T} indicator\{X_i\ is\ missing\}.$$

$$Percentile\ p_i = 100(i - 0.5) for\ (X_1', X_2', \ldots, X_T'), where\ X_1' \leq \cdots \leq X_T'\ and\ \{X_1', X_2', \ldots, X_T'\} = \{X_1, X_2, \ldots, X_T\}.$$

Computing these profiling statistics for every metric gives us a concrete overview on the quality of the merged data. For example, the number of missing values cannot be greater than or equal to number of time stamps; otherwise this attribute is merely noise. Another example could be that the clock-per-instruction (CPI) cannot be lower than 0.25, and one could use the minimum and percentile to verify. Or the CPU utilization cannot be over 100%, which can be checked using these profiling statistics. Again, we note that these rules are not the gold standards, because real cloud performance data may have violated some rules, but the data itself is correctly measured. However, the profiling statistics of the performance metrics alert the practitioner if too many inconsistent or counterintuitive phenomena occur.

### 4.6 Assess Performance Analytics Quality

Discrepancies in data processing may arise and it is difficult to spot errors, especially when the data is high dimensional. Recall in the tale of two engineers, where they write scripts respectively and independently to process the data, and then compare their results entry by entry to ensure quality. Even though this is an idealized scenario, this particular process gives an extra check on the quality of data processing.

Our tool indeed embraces this piece of philosophy of two engineers working on the same data processing problems in parallel and detecting differences along the way. Our software implements two independent scripts in parallel, using different packages in R or Python, to process the same cloud performance data. Then the software compares the processed performance datasets from two parallel channels. The comparison is done not only entry-by-entry, but also ensures the attribute names and orders are consistent.

Additionally we added a third-party involvement feature to our software. This feature enables the software user to upload his/her own data processing scripts, and compares the processed dataset from their scripts with the processed datasets using the two already implemented and parallel processing channels. This feature provides feedback on the quality of the user's processing script, and double checks the quality of the data processing scheme.

## 5 Summary and Discussion

Cloud computing has transformed the IT industry. Analysis on cloud performance has an immerse impact on the cloud computing environment, and sensible analysis can lead to better and more efficient cloud computing environment. However, cloud performance data comes in very raw form, which means they are not readily for further data analytics. In this paper, we propose a software that transforms the raw data into conventional data formats, and explain the software's several layers of quality assessment to ensure the data processing quality. We present a case study of cloud performance data, and demonstrate the effectiveness of our software at data processing and particularly the quality of data processing. We have established a methodology to evaluate and improve the quality of the analytics used for cloud performance assessment.

## Acknowledgement

The authors would like to thank the referees Dave Patterson and Emily Ren for their insightful comments and suggestions.